\documentclass{article}
\usepackage{amsmath,graphicx,mlspconf}

\usepackage{cite}
\usepackage{multirow}
\usepackage{amsfonts}

\title{ADAPTIVE APPROACH FOR SPARSE REPRESENTATIONS USING THE LOCALLY COMPETITIVE ALGORITHM FOR AUDIO}
\name{Soufiyan Bahadi,
      Jean Rouat,
      and \'{E}ric Plourde}
\address{
    NECOTIS Research Lab, Universit\'{e} de Sherbrooke, QC, Canada\\
    \{soufiyan.bahadi, jean.rouat, eric.plourde\}@usherbrooke.ca}
\begin{document}

\maketitle

\begin{abstract}
Gammachirp filterbank has been used to approximate the cochlea in sparse coding algorithms. An oriented grid search optimization was applied to adapt the gammachirp’s parameters and improve the Matching Pursuit (MP) algorithm’s sparsity along with the reconstruction quality. However, this combination of a greedy algorithm with a grid search at each iteration is computationally demanding and not suitable for real-time applications. This paper presents an adaptive approach to optimize the gammachirp’s parameters but in the context of the Locally Competitive Algorithm (LCA) that requires much fewer computations than MP. The proposed method consists of taking advantage of the LCA’s neural architecture to automatically adapt the gammachirp’s filterbank using the backpropagation algorithm. Results demonstrate an improvement in the LCA's performance with our approach in terms of sparsity, reconstruction quality, and convergence time. This approach can yield a significant advantage over existing approaches for real-time applications.
\end{abstract}
\section{Introduction}
\label{sec:intro}

Sparse code generally refers to a representation where a small number of elements from a potentially over-complete dictionary are chosen to approximate a signal. It usually generates shift-invariant representations of a given input signal with good preservation of transients and other non-stationary elements \cite{adaptive-MP}. Most of the proposed sparse coding algorithms use a greedy approach such as matching pursuit (MP) or one of its derivatives \cite{matching_pursuit, ROMP, CoSaMP, stOMP, saMP}. However, these approaches are very difficult to implement on parallel hardware. More recently, sparse code generators based on neural circuitry have emerged in the literature (\cite{LCA, MP_SNN, SP, cortical, PLCA, CLCA} among others). These neural-based architectures are much easier to implement and less computationally demanding than greedy methods.

One such approach is the Locally Competitive Algorithm (LCA)\cite{LCA}. It encodes/decodes a given signal with the smallest number of active neurons possible thanks to the lateral inhibition. It was initially developed for image and video processing using Gabor filters, then for audio signals using gammatone/compressive gammachirp filters \cite{PLCA, CLCA}. The parameters of these filters are fitted to the simultaneous noise masking data\cite{Gammachirp} which are composed of sinusoidal signals in the presence of noise. However, this dataset does not represent all types of audio data like speech, music, etc. Therefore, compressive gammachirp filters need to be adapted depending on the acoustical environment.

In that context, we propose an adaptive version of LCA that optimizes the gammachirp's filterbank. Adapting the gammachirp function was proposed by Pichevar \textit{et al.} \cite{adaptive-MP} in the context of MP using a suboptimal grid search \cite{sub-optimal}. Because of the high computational demand of a standard grid search, the frequency modulation parameter—also called the chirp parameter—was prioritized over other parameters. However, the optimization is still quite demanding as the search is performed at each iteration. While being inspired by \cite{adaptive-MP}, the proposed approach is novel in many aspects. First, the gammachirp adaptation was never used before in the context of LCA. Second, the approach benefits from the neural architecture of the LCA by adapting the gammachirp's parameters with the error back-propagation algorithm. The filterbank is therefore designed according to the acoustical environment in which the LCA would be used. This is less computationally demanding than the optimization in \cite{adaptive-MP} in the sense that the filterbank is optimized before being implemented in an application and does not need any adaptation while encoding the audio signals. Finally, we characterized each channel of the filterbank with its own parameter values, unlike previous work which used the same value of gammachirp's parameters for all filters composing the over-complete dictionary. This configuration can be compared to what is observed in biology as outer hair cells—responsible for a modulation phenomenon in the cochlea—not only have different lengths according to the coding place of the frequency on the cochlea within and across species \cite{OHC-length}, but also change dynamically their stiffness. We accordingly changed the gammachirp's filterbank properties by applying a gradient-based optimization. We present results that improve the LCA performance for audio signals using the adapted gammachrip (aGC) filterbank in comparison with the gammatone (GT) and compressive gammachirp (cGC) filterbanks.

\section{METHODS}
\label{sec:method}

\subsection{Locally competitive algorithm}
\label{ssec:lca}

The goal of LCA \cite{LCA} is to represent input signals as a linear combination of a family of atoms $\boldsymbol{D} = (\boldsymbol{\phi}_m)_{1 \leq m \leq N}$, called dictionary, where most of the coefficients $\boldsymbol{a} = (a_m)_{1 \leq m \leq N}$ are zero.
\begin{equation}
    \label{eq:recons}
    \displaystyle
    \boldsymbol{\hat{s}} = \sum_{i=1}^{N} a_{m} \boldsymbol{\phi_m} = \boldsymbol{D}\boldsymbol{a},
\end{equation}
where $\boldsymbol{\hat{s}}$ is the approximation of the input signal $\boldsymbol{s}$ and $N$ is the number of atoms. For this purpose, a recurrent neural network incorporating lateral inhibition is defined with an objective function to be minimized. This function is referred to as an energy function $E$ defined as a combination of the Mean Squared Error (MSE) between $\boldsymbol{s}$ and $\boldsymbol{\hat{s}}$, a sparsity cost penalty $S$ evaluated from the activation of neurons that corresponds to the coefficients $(a_m)_{1 \leq m \leq N}$ in (\ref{eq:recons}), and a trade-off parameter $\lambda$.
\begin{equation}
    \label{eq:energy}
    \displaystyle
    E = \frac{1}{2}||\boldsymbol{\hat{s}}-\boldsymbol{s}||^2 + \lambda S(\boldsymbol{a}).
\end{equation}
The neural dynamics are governed by the vectorized ordinary differential equation:
\begin{equation}
    \label{eq:ODE}
    \displaystyle
    \tau \frac{d\boldsymbol{u}}{dt} = \boldsymbol{p} - \boldsymbol{u} - (\boldsymbol{D}^T\boldsymbol{D}-\boldsymbol{I})\boldsymbol{a},
\end{equation}
where $\tau$ is the time constant of each neuron, $\boldsymbol{p}$ is the input signal projection on the dictionary, i.e., $\boldsymbol{p} = \boldsymbol{D}^T \boldsymbol{s}$, $\boldsymbol{u}$ is the membrane potential vector, and $\boldsymbol{I}$ is the identity matrix. Basically, the evolution of $\boldsymbol{u}$ over time depends on the input intensity $\boldsymbol{p}$ and on $-\boldsymbol{u}$ which makes these neurons behave like leaky integrators. Membrane potentials exceeding the threshold $\lambda$ produce activations whereby each activated neuron inhibits all others through horizontal connections $\boldsymbol{D}^T\boldsymbol{D}-\boldsymbol{I}$. The activation is a non-linearity $T_\lambda$ that can be sigmoidal or the hard thresholding function applied on each element $u_m$ of the potentials $\boldsymbol{u}$ such as:
\begin{equation}
    \label{eq:thresh}
    \displaystyle
    a_m = T_\lambda(u_m) =
    \begin{cases}
      0 & \text{if $|u_m| < \lambda$}\\
      u_m & \text{otherwise}
    \end{cases}.
\end{equation}
It has been shown \cite{LCA} that by imposing the following relation between activations, potentials, and the sparsity cost $S$,
\begin{equation}
    \label{eq:cost}
    \displaystyle
    \lambda\frac{\partial S(\boldsymbol{a})}{\partial a_m} = u_m - a_m
\end{equation}
the evolution over time of $\boldsymbol{u}$ becomes negatively proportional to the derivative of the energy $E$ with respect to $\boldsymbol{a}$, (\ref{eq:energy}) is therefore minimized.
Using the relation (\ref{eq:cost}), $S$ can be  $l^1$ norm or $l^0$-like norm depending on the type of the activation function $T_\lambda$ \cite{LCA}.
\begin{table}[t]
\centering
\begin{tabular}{c|c|c|}
\cline{2-3}
\multicolumn{1}{l|}{}                             & Parameter & Value \\ \hline
\multicolumn{1}{|c|}{\multirow{3}{*}{Dictionary}} & k         & 16    \\
\multicolumn{1}{|c|}{}                            & $F_l$     & 1024  \\
\multicolumn{1}{|c|}{}                            & r         & 10    \\ \hline
\multicolumn{1}{|c|}{\multirow{3}{*}{GT}}         & c         & 0     \\
\multicolumn{1}{|c|}{}                            & b         & 1     \\
\multicolumn{1}{|c|}{}                            & l         & 4     \\ \hline
\multicolumn{1}{|c|}{\multirow{3}{*}{cGC}}        & c         & 0.979 \\
\multicolumn{1}{|c|}{}                            & b         & 1.14  \\
\multicolumn{1}{|c|}{}                            & l         & 4     \\ \hline
\end{tabular}
    \caption{Hyper-parameters of the Dictionary, the GT, and the cGC.}
    \label{tab:HP}
\end{table}

\subsection{Dictionary}
\label{sec:dict}

As defined in \cite{Gammachirp}, a gammachirp filter impulse response is a monotonically frequency-modulated carrier—a chirp—with an envelope that is a gamma distribution function,
\begin{equation}
    \label{eq:gam}
    \displaystyle
    g_i(t) = t^{l-1} e^{-2 \pi b \text{ERB}(f_i)t} \cos(2 \pi f_i t + c \ln(t)),
\end{equation}
where $l$ and $b$ are gamma distribution parameters that control the attack and the decay of the kernel, $c$ is referred to as the chirp parameter which modulates the carrier frequency allowing to slightly modify the instantaneous frequency and $f_i$ is the central frequency. $\text{ERB}(.)$ is a linear transformation of $f_i$ on the Equivalent Rectangular Bandwidth scale \cite{glasberg}:
\begin{equation}
    \label{ERB}
    \displaystyle
    \text{ERB}(f_i) = 24.7 + 0.108 f_i.
\end{equation}
Inspired by the fact that individual channel adaptation occurs in the cochlea through the outer hair cells, we characterize each channel with its own parameters  $l_i$, $b_i$ and $c_i$ instead of sharing the same values among all filters. Hence, (\ref{eq:gam}) becomes:
\begin{equation}
    \label{eq:gam2}
    \displaystyle
    g_i(t) = t^{l_i-1} e^{-2 \pi b_i \text{ERB}(f_i)t} \cos(2 \pi f_i t + c_i \ln(t)).
\end{equation}
With this equation we create a discrete time dictionary by striding each filter across the sampled signal,
\begin{equation}
    \label{eq:dict}
    \displaystyle
    \boldsymbol{D}^T =
    \begin{bmatrix}
    g_0[0] & g_0[1] & ... & 0 & 0 & ...\\
    \boldsymbol{0_{1 \times r}} & g_0[0] & g_0[1] & ... & 0 & ...\\
    \boldsymbol{0_{1 \times r}} & \boldsymbol{0_{1 \times r}} & g_0[0] & g_0[1] & ... & ...\\
    \vdots & & & & & & \vdots \\
    g_i[0] & g_i[1] & ... & 0 & 0 & ...\\
    \boldsymbol{0_{1 \times r}} & g_i[0] & g_i[1] & ... & 0 & ...\\
    \boldsymbol{0_{1 \times r}} & \boldsymbol{0_{1 \times r}} & g_i[0] & g_i[1] & ... & ...\\
    \vdots & & & & & & \vdots \\
    g_k[0] & g_k[1] & ... & 0 & 0 & ...\\
    \boldsymbol{0_{1 \times r}} & g_k[0] & g_k[1] & ... & 0 & ...\\
    \boldsymbol{0_{1 \times r}} & \boldsymbol{0_{1 \times r}} & g_k[0] & g_k[1] & ... & ...\\
    \end{bmatrix},
\end{equation}
where $\boldsymbol{0_{1 \times r}}$ is a row vector of zeros of length $r$ (stride size) and $k$ is the number of channels. The values used for these parameters and for the length of the sampled impulse response of each filter $F_l$ are shown in Tab.~\ref{tab:HP}.

\begin{table}[t]
    \centering
    \begin{tabular}{ |c|c| }
     \hline
     Parameter & Value \\ 
     \hline
     $\tau$ & $0.01$ \\
     $\Delta t$ & $0.0001$ \\
     Iters & $64$ \\
     \hline
    \end{tabular}
    \caption{Hyper-parameters of the LCA. $\tau$ is the time constant of neurons, $\Delta t$ is the step size of Euler's method, and Iters is the number of iterations of LCA. }
    \label{tab:lcaHP}
\end{table}

\subsection{Gradient-based optimization of chirp parameters}
\label{ssec:grad}

We implemented backpropagation through time (BPTT) for the adaptation of the gammachirp filterbank. In order to limit memory usage, truncated BPTT \cite{TBPTT} was used instead of the standard BPTT along with a buffer in which losses and steady states of the LCA are stacked. Once this buffer is full, gradients of all losses are back-propagated.

Through the differentiable objective function (\ref{eq:energy}), $E$ depends on the dictionary $\boldsymbol{D}$ and the activation of neurons $\boldsymbol{a}$ which is a differentiable non linearity applied to membrane potentials $\boldsymbol{u}$ as given in (\ref{eq:thresh}). Futhermore, (\ref{eq:ODE}) shows the differentiable relationship between $\boldsymbol{u}$'s dynamics and the dictionary $\boldsymbol{D}$ which depends in its turn, through the differentiable gammachirp function (\ref{eq:gam2}), on $\boldsymbol{c}$, $\boldsymbol{b}$ and $\boldsymbol{l}$. By assuming that $\boldsymbol{u}$ depends on $\boldsymbol{D}$ and differentiable with respect to it as it is the case for $\frac{d\boldsymbol{u}}{dt}$, the gradients can be computed using the chain rule,
\begin{align}
    \label{eq:chain}
    \displaystyle
    \begin{aligned}
        & \frac{\partial E}{\partial (\boldsymbol{c}|\boldsymbol{b}|\boldsymbol{l})} = \frac{\partial \frac{1}{2}||\boldsymbol{D}\boldsymbol{a} - \boldsymbol{s}||^2}{\partial \boldsymbol{D}\boldsymbol{a}} \frac{\partial \boldsymbol{D}\boldsymbol{a}}{\partial \boldsymbol{D}} \frac{\partial \boldsymbol{D}}{\partial (\boldsymbol{c}|\boldsymbol{b}|\boldsymbol{l})}\\ & \qquad\qquad + \frac{d \boldsymbol{S(\boldsymbol{a})}}{d \boldsymbol{a}} \frac{d \boldsymbol{a}}{d \boldsymbol{u}} \frac{\partial \boldsymbol{u}}{\partial \boldsymbol{D}} \frac{\partial \boldsymbol{D}}{\partial (\boldsymbol{c}|\boldsymbol{b}|\boldsymbol{l})}
    \end{aligned},
\end{align}
\begin{figure}[htb]
\begin{minipage}[b]{1.0\linewidth}
  \centering
  \centerline{\includegraphics[width=8.5cm]{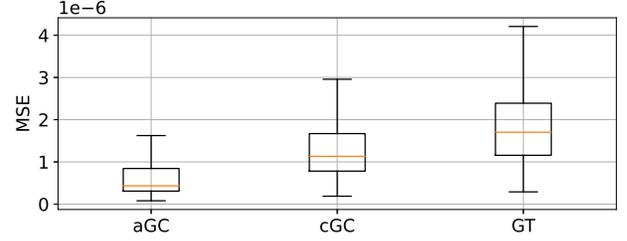}}
  \centerline{(a) MSE: the residual energy divided by the signal length}\medskip
\end{minipage}
\begin{minipage}[b]{1.0\linewidth}
  \centering
  \centerline{\includegraphics[width=8.5cm]{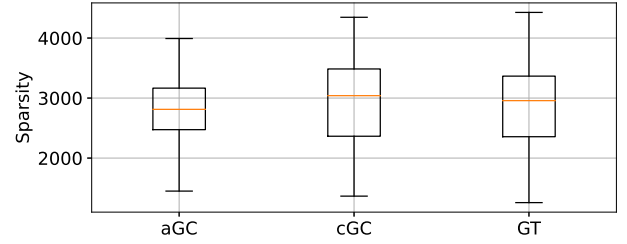}}
  \centerline{(b) Sparsity: number of active neurons}\medskip
\end{minipage}
\caption{(a) Quality and (b) sparsity distributions on the test set for the LCA with dictionaries generated by the adapted Gammachip (aGC), compressive Gammachirps (cGC) and Gammatone (GT). After optimization on the training set, we evaluated the test set by running LCA for 64 iterations. Results are illustrated by boxplots where the median is represented by the line inside each box, the bottom line of the box corresponds to the first quartile Q1, the box top line corresponds to the third quartile Q3 and the bottom and top horizontal lines represent respectively the lowest and the highest data points.}
\label{fig:dist}
\end{figure}
where $|$ is the "OR" operator and $\boldsymbol{S(\boldsymbol{a})} = \sum_{m=1}^N S(a_m)$. To validate our assumption on $\boldsymbol{u}$'s differentiability, we need to compute $\boldsymbol{u}(t)$. In fact, (\ref{eq:ODE}) leads to piece wise exponential traces for the membrane potential which is computed by Euler's method,
\begin{align}
    \label{eq:euler}
    \displaystyle
    \begin{aligned}
        & \boldsymbol{u}(t) = \frac{\Delta t}{\tau} [\boldsymbol{D}^T \boldsymbol{s} - (\boldsymbol{D}^T\boldsymbol{D}-\boldsymbol{I}) \boldsymbol{a}(t-\Delta t)]\\ & \qquad\qquad+(1-\frac{\Delta t}{\tau})\boldsymbol{u}(t - \Delta t)
    \end{aligned},
\end{align}
which shows that $\boldsymbol{u(t)}$ is differentiable with respect to $\boldsymbol{D}$.

Moreover, the first two factors in the second term of (\ref{eq:chain})—which correspond to the sparsity error gradient with respect to $\boldsymbol{u}$—are a particular case study when it comes to using hard thresholding function (\ref{eq:thresh}). Using this activation function, the gradient of the cost penalty $S$ as described in (\ref{eq:cost}) becomes for each neuron $m$:
\begin{equation}
    \label{eq:12}
    \displaystyle
    \lambda \frac{\partial S(\boldsymbol{a})}{\partial a_m} = 
    \begin{cases}
      u_m & \text{if $|u_m| < \lambda$}\\
      0 & \text{otherwise}
    \end{cases}.
\end{equation}
Since the derivative of the hard thresholding function with respect to $u_m$ is,
\begin{equation}
    \label{eq:13}
    \displaystyle
    \frac{d a_m}{d u_m} = T'_{\lambda}(u_m) = 
    \begin{cases}
      0 & \text{if $|u_m| < \lambda$}\\
      1 & \text{otherwise}
    \end{cases},
\end{equation}
the gradient of the second term of $E$ with respect to $u_m$, which is computed by multiplying (\ref{eq:12}) and (\ref{eq:13}), becomes null for all $u_m$ values. In other words, the gradient of the sparsity cost is always canceled in the back-propagation. Consequently, $\frac{\partial a_m(t)}{\partial u_m(t)}$ should be ignored for all neurons to preserve the sparsity gradient flow. We therefore set $\frac{\partial \boldsymbol{a}(t)}{\partial \boldsymbol{u}(t)}$ to $1$. This modification gives a new gradient of $E$ which is proportional to the exact one because $T_{\lambda}$ is a monotonically increasing function. Taking into account the above, (\ref{eq:chain}) becomes,

\begin{align}
    \label{eq:dE/dc}
    \displaystyle
    \begin{aligned}
        & \frac{\partial E}{\partial (\boldsymbol{c}|\boldsymbol{b}|\boldsymbol{l})} = \frac{\partial \frac{1}{2}||\boldsymbol{D}\boldsymbol{a} - \boldsymbol{s}||^2}{\partial \boldsymbol{D}\boldsymbol{a}} \frac{\partial \boldsymbol{D}\boldsymbol{a}}{\partial \boldsymbol{D}} \frac{\partial \boldsymbol{D}}{\partial (\boldsymbol{c}|\boldsymbol{b}|\boldsymbol{l})}\\ & \qquad\qquad + \frac{d \boldsymbol{S(\boldsymbol{a})}}{d \boldsymbol{a}} \frac{\partial \boldsymbol{u}}{\partial \boldsymbol{D}} \frac{\partial \boldsymbol{D}}{\partial (\boldsymbol{c}|\boldsymbol{b}|\boldsymbol{l})}
    \end{aligned},
\end{align}
In this work, we used hard thresholding as an activation function because it leads to better convergence of LCA \cite{LCA}. Accordingly, we used the modified gradient from equation (\ref{eq:dE/dc}). We also computed all gradients using computation graphs of the Pytorch library\cite{pytorch}.

\begin{figure}[t]
\begin{minipage}[b]{1.0\linewidth}
  \centering
  \centerline{\includegraphics[width=8.5cm]{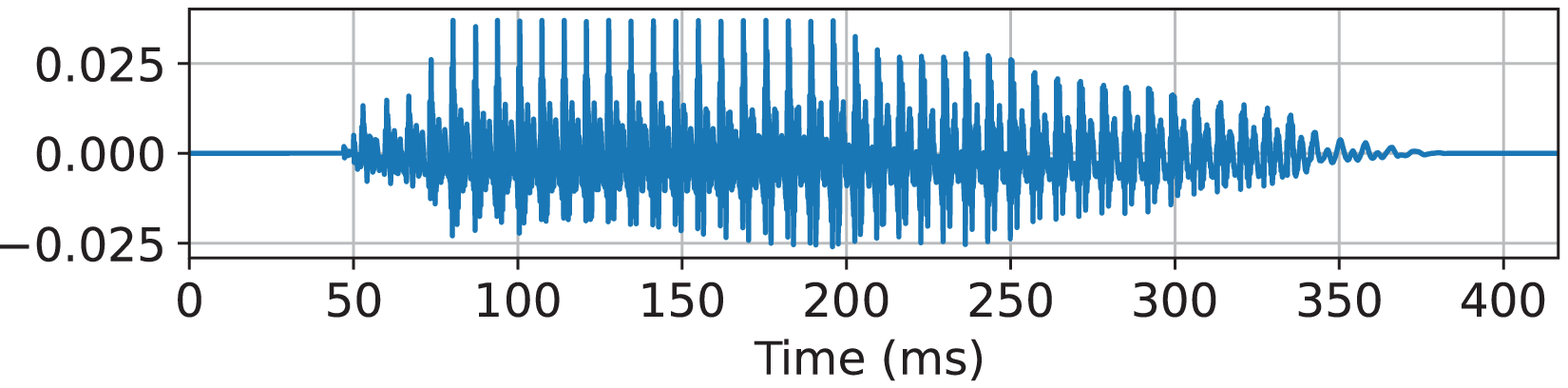}}
  \centerline{(a) Waveform: "eight"}\medskip
\end{minipage}
\begin{minipage}[b]{1.0\linewidth}
  \centering
  \centerline{\includegraphics[width=8.5cm]{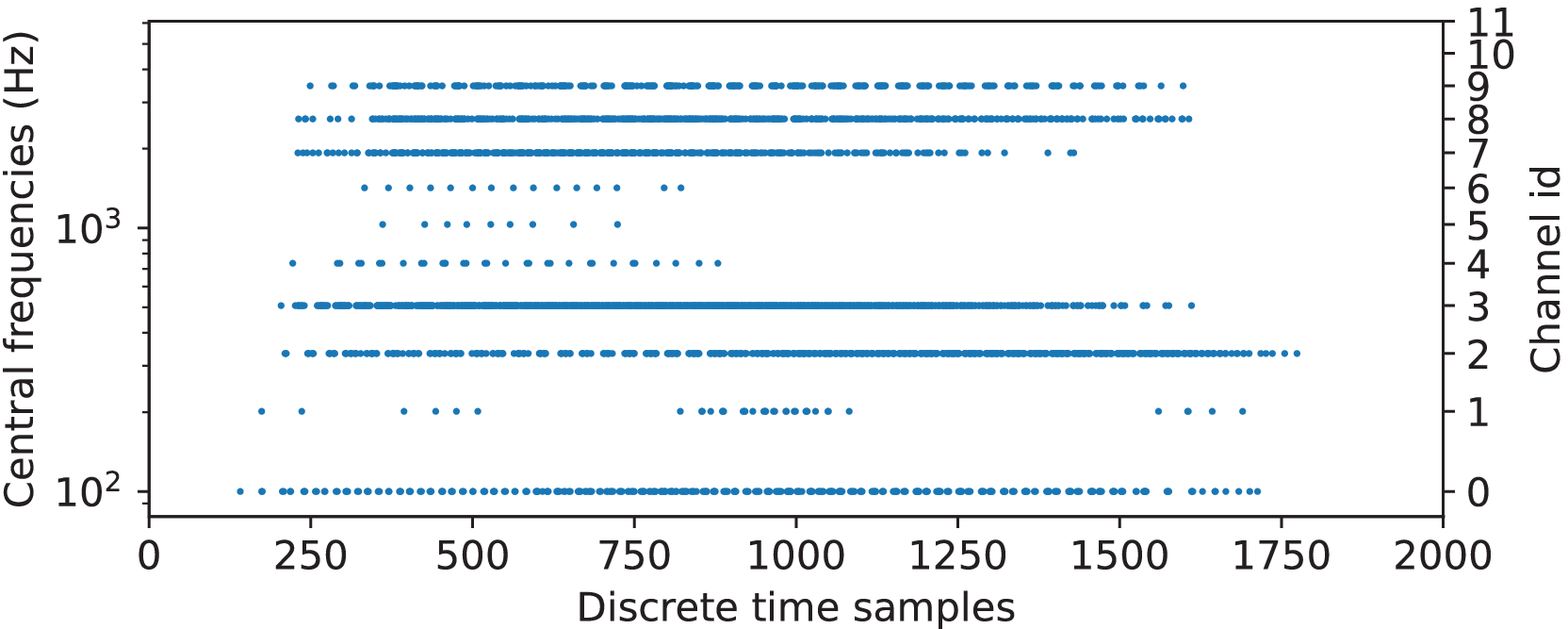}}
  \centerline{(b) Spikegram using GT: 2970 spikes}\medskip
\end{minipage}
\begin{minipage}[b]{1.0\linewidth}
  \centering
  \centerline{\includegraphics[width=8.5cm]{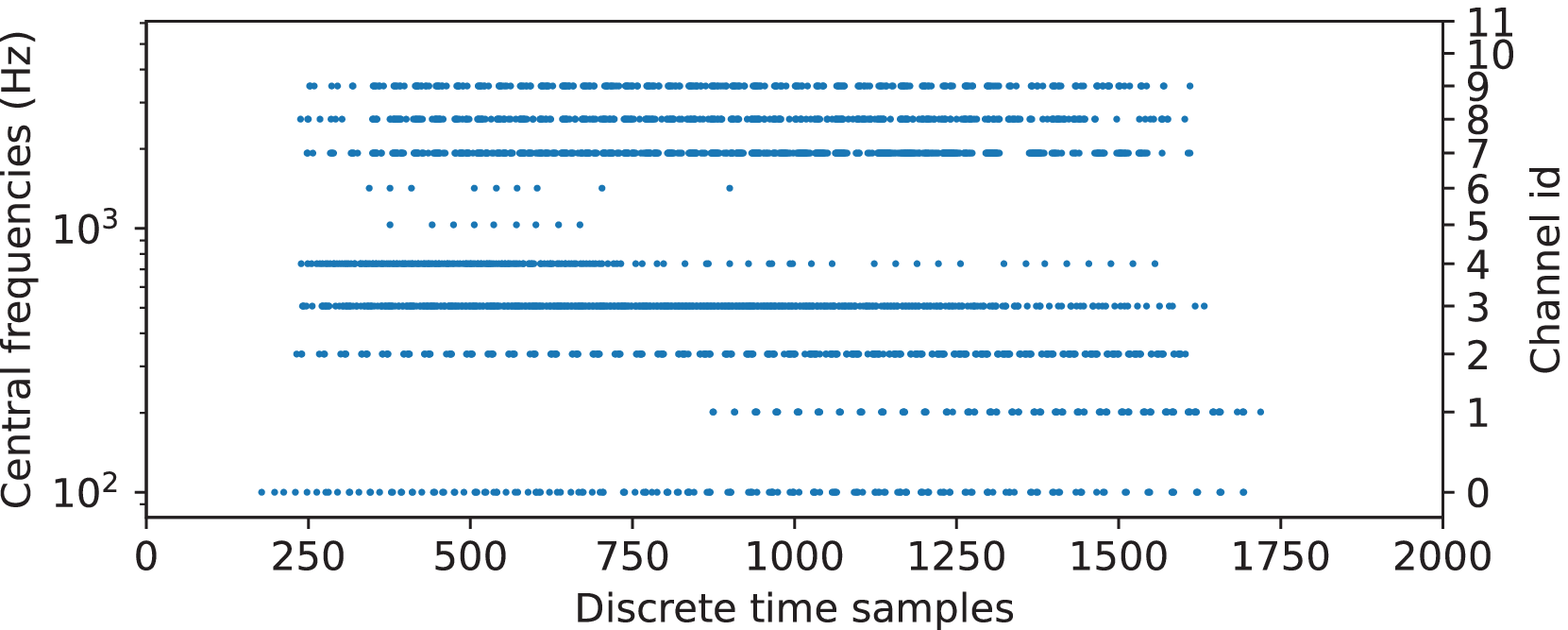}}
  \centerline{(c) Spikegram using aGC: 2694 spikes}\medskip
\end{minipage}
\caption{(a) Pronounced "eight" waveform and spikegrams using (b) GT and (c) aGC dictionaries. Spikegrams were generated from LCA output coefficients by creating a spike (blue dot) at the corresponding channel and discrete time. For the sake of simplicity, the amplitudes of spikes are not represented and channels from 11 to 15 are not plotted since they did not spike.}
\label{fig:spg}
\end{figure}

\subsection{Hyper-parameters and data set}
\label{sec:hyper}

We initialized the learning with a dictionary composed of GT atoms as defined in \cite{glasberg}. The parameters of GT are indicated in Tab.~\ref{tab:HP}. We used the Adam optimizer with a learning rate of $0.0002$ and a mini-batch size of $8$. The number of epochs is $10$ and the size of the buffer used for TBTT is $8$. The parameters of cGC Tab.~\ref{tab:HP} are specified in \cite{Park} and inspired by Irino and Patterson \cite{ Gammachirp}.

For this study \footnote{https://github.com/SoufiyanBAHADI/ALCA}, we are interested in audio signals of spoken numbers. We chose the Heidelberg data set \cite{Heidelberg} which consists of approximately $10 000$ recordings of spoken digits from zero to nine in both English and German languages including 12 speakers in total. We used only the English set composed of 4011 recordings for training and 1079 recordings for testing. This data set is optimized for recording quality and precise audio alignment and it is not released under a proprietary license.

\begin{figure}[t]
\begin{minipage}[b]{1.0\linewidth}
  \centering
  \centerline{\includegraphics[width=8.5cm]{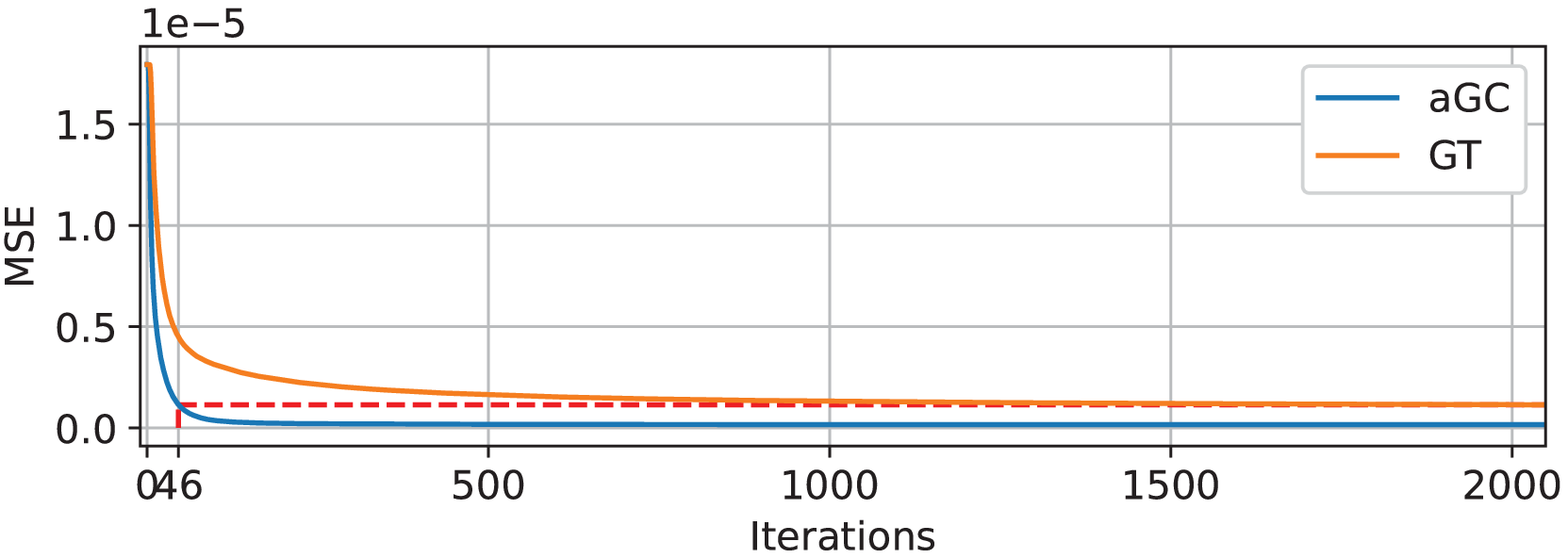}}
  \centerline{(a)}\medskip
\end{minipage}
\begin{minipage}[b]{1.0\linewidth}
  \centering
  \centerline{\includegraphics[width=8.5cm]{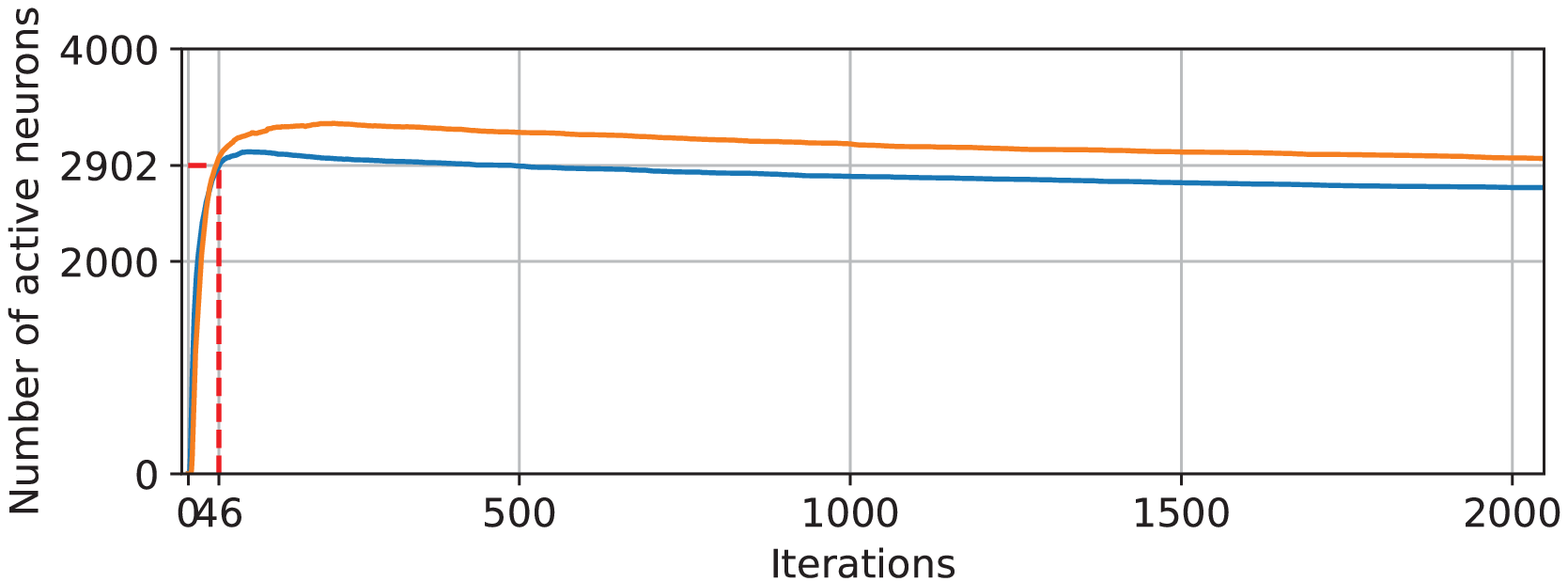}}
  \centerline{(b)}\medskip
\end{minipage}
\caption{The evolution of the (a) MSE and the (b) number of active neurons while LCA is processing the digit "eight" of Fig.~\ref{fig:spg}. The red dashed line shows the iteration corresponding to the minimal MSE in (a) and the number of active neurons corresponding to that iteration in (b). }
  \label{fig:losses}
\end{figure}

\section{RESULTS AND DISCUSSION}
\label{sec:rad}

\begin{figure}[t]
\begin{minipage}[b]{1.0\linewidth}
  \centering
  \centerline{\includegraphics[width=8.5cm]{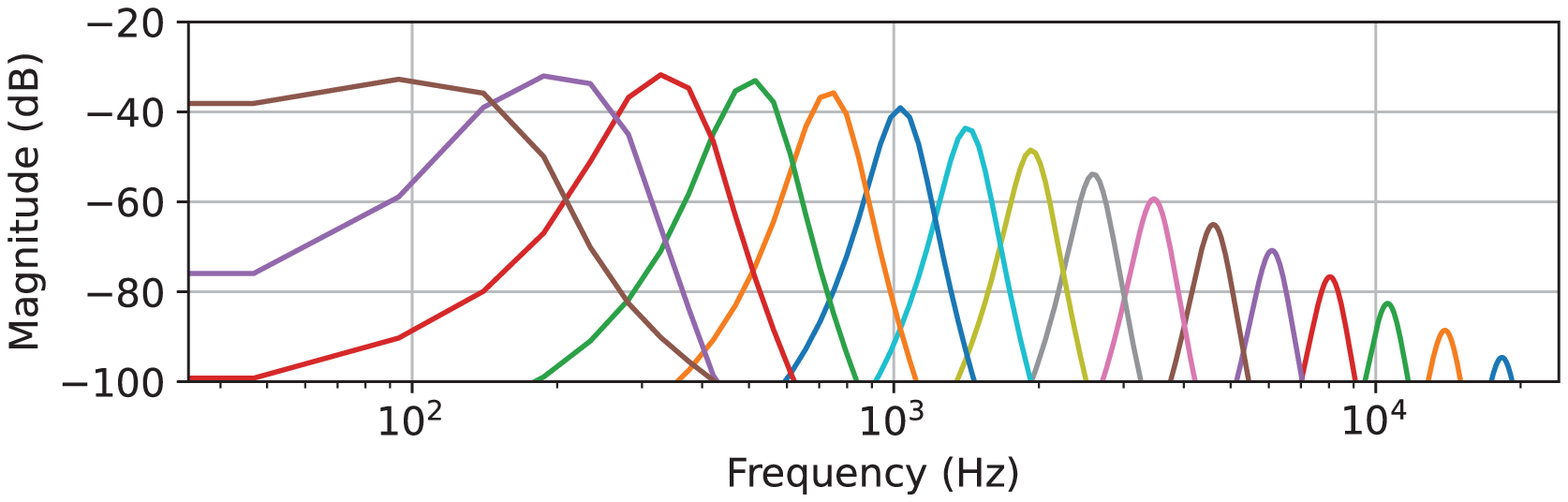}}
  \centerline{(a) Original filterbank: GT}\medskip
\end{minipage}
\begin{minipage}[b]{1.0\linewidth}
  \centering
  \centerline{\includegraphics[width=8.5cm]{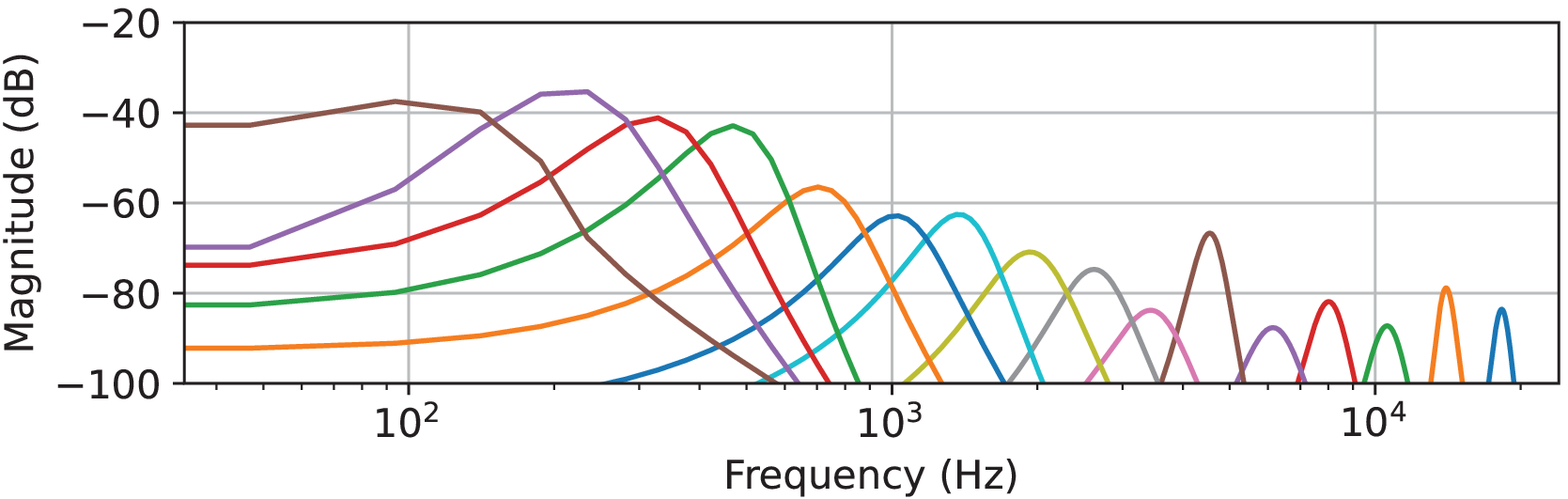}}
  \centerline{(b) Filterbank after optimization: aGC}\medskip
\end{minipage}
  \caption{Magnitude in dB of (a) GT and (b) aGC filterbanks vs frequency in logarithmic scale. The energy of the impulse responses of all filters is normalized.}
  \label{fig:FB}
\end{figure}

\begin{figure}[t]
\begin{minipage}[b]{.48\linewidth}
  \centering
  \centerline{\includegraphics[width=4.0cm]{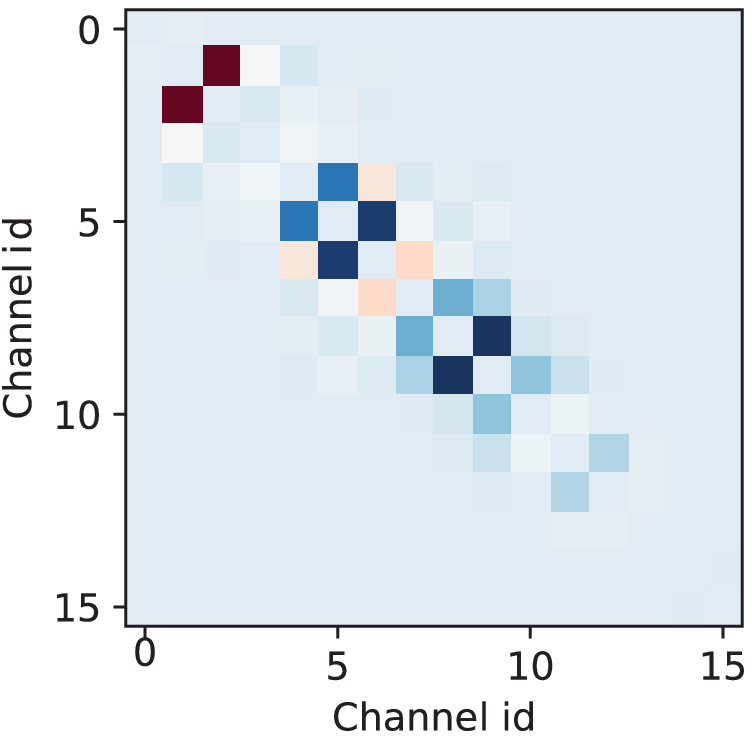}}
  \centerline{(a) aGC filterbank}\medskip
\end{minipage}
\hfill
\begin{minipage}[b]{0.48\linewidth}
  \centering
  \centerline{\includegraphics[width=4.0cm]{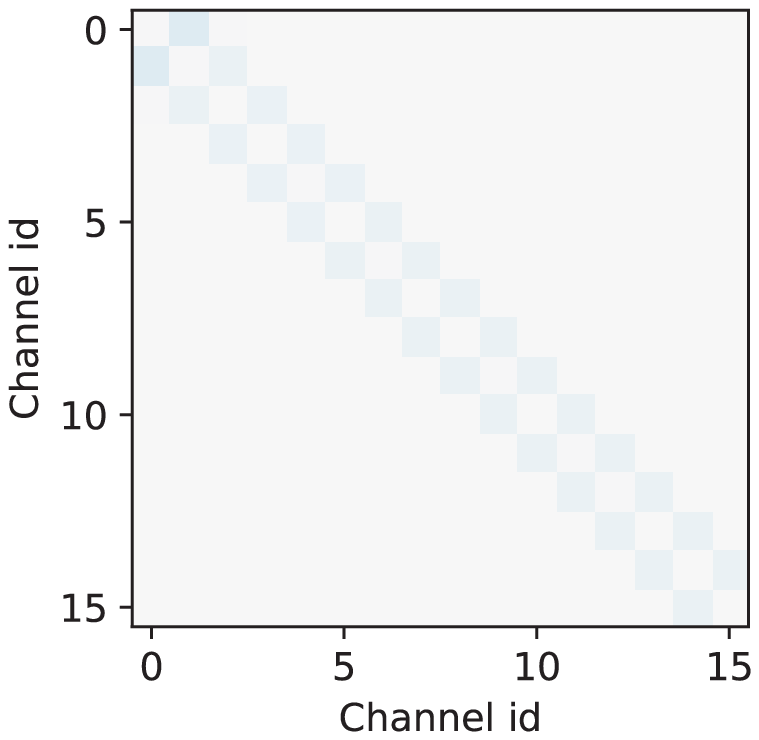}}
  \centerline{(b) GT filterbank}\medskip
\end{minipage}
\begin{minipage}[b]{1.0\linewidth}
  \centering
  \centerline{\includegraphics[width=8.5cm]{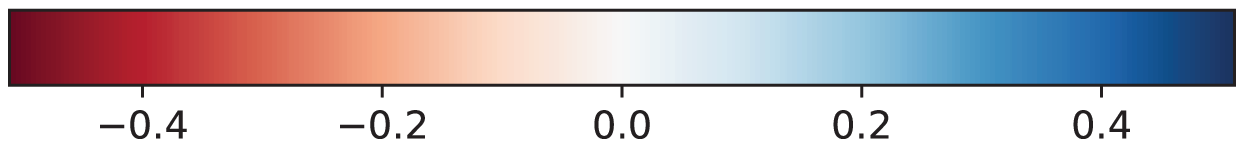}}\medskip
\end{minipage}
  \caption{Inhibition weights between channels of (a) aGC and (b) GT filterbanks.}
  \label{fig:R}
\end{figure}

After running the LCA with the parameters in Tab.~\ref{tab:lcaHP} using the GT, cGC, and aGC dictionaries, we show in Fig.~\ref{fig:dist} the distribution of the MSE and spike count on the test set recordings. With the aGC dictionary, the LCA achieved the lowest median MSE and the smallest interquartile range (Fig.~\ref{fig:dist}a). Even though the cGC led to the second-best reconstruction quality, only 25\% of reconstructed recordings using this dictionary are as good as 75\% of those reconstructed with the aGC. In addition, the aGC dictionary is able to represent signals with the least median number of active neurons and the narrowest interquartile (Fig.~\ref{fig:dist}b). Overall, the proposed adaptation process allowed to represent recordings in the sparsest way and with the best quality.

Fig.~\ref{fig:spg}a illustrates an example of a pronounced digit from the test set with its "spikegrams" generated using the GT (Fig.~\ref{fig:spg}b) and aGC (Fig.~\ref{fig:spg}c) dictionaries. Each dot corresponds to a coefficient—referred to as a spike—in the sparse approximation. For the sake of clarity, we did not illustrate the amplitude of each spike. The adaptation of the filterbank does more than just a compression of the representation. It adjusts GT filterbank properties to find filters that better represent the signal with fewer spikes. For example, the channel 4—which corresponds to the central frequency of $734.5$ Hz—shows very few spikes in the discrete time window $[250, 750]$ (Fig.~\ref{fig:spg}b). However, the same channel using aGC (Fig.~\ref{fig:spg}c) shows more spikes in the same time window although aGC reduced spikes number by more than 9\% (from 2970 spikes to 2694 spikes).

To evaluate the impact of adaptation on LCA convergence, we ran the LCA for 2048 iterations on the same example used above with the aGC and GT dictionaries and plotted the evolution of the MSE and the number of spikes in Fig.~\ref{fig:losses}. Using aGC, LCA achieved at iteration $46$ the quality and sparsity that it achieved at the last iteration (i.e., $2047$) using the GT. Thus, the LCA converged about $43.6$ times faster with the aGC than with the GT. This can be explained by the fact that the LCA's objective function (\ref{eq:energy}) is minimized by two processes: the membrane potential dynamics and the adaptation of the dictionary properties. By visualizing the aGC and GT filterbanks Fig.~\ref{fig:FB}, we observe that after optimization the aGC filters are less selective with wider bandwidths for low frequencies than in their initial state (the GT filters). This makes the filter's overlap larger in the frequency domain which increases their similarities. Therefore, the absolute values of lateral inhibition weights between channels—referred to as spectral inhibition weights—become larger as shown in Fig.~\ref{fig:R}. We notice that the aGC's spectral inhibition weights (Fig.~\ref{fig:R}a) have significant values, whereas for GT all spectral inhibition weights (Fig.~\ref{fig:R}b) are null except between adjacent channels. Accordingly, after optimization, new interactions were established between neurons representing channels that are not directly adjacent, and local competition is thereby enhanced. It is, therefore, more difficult for a neuron to be activated unless its receptive field fits well the input and no neuron with a very similar receptive field is already activated.

\section{CONCLUSION}
\label{sec:conclusion}

Preliminary results demonstrate that the LCA can achieve relatively better sparsity and reconstruction quality when using the proposed gammachirp filterbank adaptation compared to predefined gammachirp as used in literature. Such adaptation also improved the LCA convergence time. The proposed approach may therefore be a good candidate to be used in a real-time application. In addition, being inspired by the fact that outer hair cells have different lengths and change dynamically their stiffness, we characterized each channel with its own parameter values. Results show that the gains and bandwidths of the filters were influenced by the optimization and induced larger overlapping regions in the frequency domain which increased the similitudes of the filters. Therefore, lateral inhibition weights became more significant causing more competition between neurons to represent the signal. In future work, we will compare the proposed approach with previous methods that utilizes the MP algorithm and we will also test the approach in the context of a real-time application to fully understand its impact on the time response.

\section{ACKNOWLEDGMENT}
\label{sec:ack}

The authors would like to thank the Fonds de recherche du Qu\'{e}bec - Nature et technologies for funding this research and NVIDIA for donating the GTX1080 and Titan Xp GPUs. We would also like to thank reviewers for their comments.

\bibliographystyle{IEEEbib}
\bibliography{strings,refs}

\begin{thebibliography}{10}

\bibitem{adaptive-MP}
R.~Pichevar, H.~Najaf-Zadeh, L.~Thibault, and H.~Lahdili,
\newblock ``Auditory-inspired sparse representation of audio signals,''
\newblock {\em Speech Communication}, vol. 53, no. 5, pp. 643--657, 2011,
\newblock Perceptual and Statistical Audition.

\bibitem{matching_pursuit}
S.G. Mallat and Z.~Zhang,
\newblock ``Matching pursuits with time-frequency dictionaries,''
\newblock {\em IEEE Transactions on Signal Processing}, vol. 41, no. 12, pp.
  3397--3415, 1993.

\bibitem{ROMP}
D.~Needell and R.~Vershynin,
\newblock ``Uniform {Uncertainty} {Principle} and {Signal} {Recovery} via
  {Regularized} {Orthogonal} {Matching} {Pursuit},''
\newblock {\em Foundations of Computational Mathematics}, vol. 9, no. 3, pp.
  317--334, June 2009.

\bibitem{CoSaMP}
D.~Needell and J.A. Tropp,
\newblock ``Cosamp: Iterative signal recovery from incomplete and inaccurate
  samples,''
\newblock {\em Applied and Computational Harmonic Analysis}, vol. 26, no. 3,
  pp. 301--321, 2009.

\bibitem{stOMP}
D.~L. Donoho, Y.~Tsaig, I.~Drori, and J.~Starck,
\newblock ``Sparse solution of underdetermined systems of linear equations by
  stagewise orthogonal matching pursuit,''
\newblock {\em IEEE Transactions on Information Theory}, vol. 58, no. 2, pp.
  1094--1121, 2012.

\bibitem{saMP}
T.T. Do, L~Gan, N~Nguyen, and T.D. Tran,
\newblock ``Sparsity adaptive matching pursuit algorithm for practical
  compressed sensing,''
\newblock in {\em Asilomar Conference on Signals, Systems and Computers}, 2008,
  pp. 581--587.

\bibitem{LCA}
C.J. Rozell, D.H. Johnson, R.G. Baraniuk, and B.A. Olshausen,
\newblock ``{Sparse Coding via Thresholding and Local Competition in Neural
  Circuits},''
\newblock {\em Neural Computation}, vol. 20, no. 10, pp. 2526--2563, 2008.

\bibitem{MP_SNN}
L.~Perrinet,
\newblock ``Efficient source detection using integrate-and-fire neurons,''
\newblock in {\em Artificial Neural Networks: Biological Inspirations}, Berlin,
  Heidelberg, 2005, pp. 167--172, Springer Berlin Heidelberg.

\bibitem{SP}
K.K. Herrity, A.C. Gilbert, and J.A. Tropp,
\newblock ``Sparse approximation via iterative thresholding,''
\newblock in {\em IEEE International Conference on Acoustics Speech and Signal
  Processing Proceedings}, 2006, vol.~3, p.~3.

\bibitem{cortical}
M.~Rehn and F.T. Sommer,
\newblock ``A network that uses few active neurones to code visual input
  predicts the diverse shapes of cortical receptive fields,''
\newblock {\em Journal of Computational Neuroscience}, vol. 22, no. 2, pp.
  135--146, Oct. 2006.

\bibitem{PLCA}
R.~Pichevar, H.~Najaf-Zadeh, and F.~Mustiere,
\newblock ``Neural-based approach to perceptual sparse coding of audio
  signals,''
\newblock in {\em International Joint Conference on Neural Networks (IJCNN)},
  2010, pp. 1--8.

\bibitem{CLCA}
A.S. Charles, A.A. Kressner, and C.J. Rozell,
\newblock ``A causal locally competitive algorithm for the sparse decomposition
  of audio signals,''
\newblock in {\em Digital Signal Processing and Signal Processing Education
  Meeting (DSP/SPE)}, 2011, pp. 265--270.

\bibitem{Gammachirp}
T.~Irino and R.D. Patterson,
\newblock ``A time-domain, level-dependent auditory filter: The gammachirp,''
\newblock {\em The Journal of the Acoustical Society of America}, vol. 101, no.
  1, pp. 412--419, 1997.

\bibitem{sub-optimal}
R.~Gribonval,
\newblock ``Fast matching pursuit with a multiscale dictionary of gaussian
  chirps,''
\newblock {\em IEEE Transactions on Signal Processing}, vol. 49, no. 5, pp.
  994--1001, 2001.

\bibitem{OHC-length}
R.~Pujol, M.~Lenoir, S.~Ladrech, F.~Tribillac, and G.~Rebillard,
\newblock ``Correlation between the length of outer hair cells and the
  frequency coding of the cochlea,''
\newblock in {\em Auditory Physiology and Perception}, Y.~CAZALS, K.~HORNER,
  and L.~DEMANY, Eds., pp. 45--52. Pergamon, 1992.

\bibitem{glasberg}
R.~D. Patterson and B.~Moore,
\newblock ``Auditory filters and excitation patterns as representations of
  frequency resolution,''
\newblock {\em Frequency selectivity in hearing}, pp. 123--177, 1986.

\bibitem{TBPTT}
H.~Jaeger,
\newblock ``Tutorial on training recurrent neural networks, covering {BPPT},
  {RTRL}, {EKF} and the echo state network approach,''
\newblock {\em GMD-Forschungszentrum Informationstechnik, 2002.}, vol. 5, 01
  2002.

\bibitem{pytorch}
A.~Paszke, S.~Gross, F.~Massa, A.~Lerer, F.~Bradbury, G.~Chanan, T.~Killeen,
  Z.~Lin, N.~Gimelshein, L.~Antiga, A.~Desmaison, A.~Kopf, E.~Yang, Z.~DeVito,
  M.~Raison, A.~Tejani, S.~Chilamkurthy, B.~Steiner, L.~Fang, J.~Bai, and
  S.~Chintala,
\newblock ``Pytorch: An imperative style, high-performance deep learning
  library,''
\newblock in {\em Advances in Neural Information Processing Systems},
  H.~Wallach, H.~Larochelle, A.~Beygelzimer, F.~d\textquotesingle
  Alch\'{e}-Buc, E.~Fox, and R.~Garnett, Eds. 2019, vol.~32, Curran Associates,
  Inc.

\bibitem{Park}
A.~Park,
\newblock ``Using the gammachirp filter for auditory analysis of speech,''
\newblock unpublished, May 2003.

\bibitem{Heidelberg}
B.~Cramer, Y.~Stradmann, J.~Schemmel, and F.~Zenke,
\newblock ``The heidelberg spiking data sets for the systematic evaluation of
  spiking neural networks,''
\newblock {\em IEEE Transactions on Neural Networks and Learning Systems}, pp.
  1--14, 2020.

\end{thebibliography}

\end{document}